\documentclass[cits]{PoS}

\usepackage[squaren,Gray,cdot]{SIunits}
\usepackage{amsmath,amssymb,braket,slashed,mathrsfs}
\usepackage[numbers,square]{natbib}
\usepackage{float}
\newcommand{\eV}{\electronvolt}
\newcommand{\fm}{\femto\meter}
\DeclareMathOperator{\bigo}{O}
\title{Non-degenerate light quark masses from 2+1f lattice QCD+QED}

\ShortTitle{Light quark masses from lattice QCD+QED}

\author{\speaker{Shane Drury}\\
		School of Physics \&\ Astronomy, University of Southampton, 
        SO17 1BJ, UK\\
        E-mail: \email{srd1g10@soton.ac.uk}}
        
\author{Tom Blum\\
		Physics Department, University of Connecticut, Storrs, CT 06269, USA and\\
		RIKEN-BNL Research Center, Brookhaven National Laboratory, Upton, NY 11973, USA\\
        E-mail: \email{tblum@phys.uconn.edu}}
        
\author{Masashi Hayakawa\\
		Department of Physics, Nagoya University, Nagoya 464-8602, Japan\\
        E-mail: \email{hayakawa@eken.phys.nagoya-u.ac.jp}}
        
\author{Taku Izubuchi\\
		Brookhaven National Laboratory, Upton, NY 11973, USA and \\
		RIKEN-BNL Research Center, Brookhaven National Laboratory, Upton, NY 11973, USA\\
        E-mail: \email{izubuchi@bnl.gov}}
        
\author{Chris Sachrajda\\
		School of Physics \&\ Astronomy, University of Southampton, 
        SO17 1BJ, UK\\
        E-mail: \email{cts@soton.ac.uk}}
        
\author{Ran Zhou\\
		Fermi National Accelerator Laboratory, Batavia, Illinois, USA and \\
		Department of Physics, Indiana University, Bloomington, IN, USA\\
        E-mail: \email{zhouran@indiana.edu}}

\abstract{We report on a calculation of the effects of isospin breaking in Lattice QCD+QED. This involves using Chiral Perturbation Theory with Electromagnetic corrections to find the renormalized, non-degenerate, light quark masses. The calculations are carried out on QCD ensembles generated by the RBC and UKQCD collaborations using Domain Wall Fermions and the Iwasaki and Iwasaki+DSDR Gauge Actions with unitary pion masses down to 170 MeV. Non-compact QED is treated in the quenched approximation. The simulations use a $32^3$ lattice size with $a^{-1}=\unit{2.28(3)}{\giga\eV}$ (Iwasaki) and 1.37(1) (Iwasaki+DSDR). This builds on previous work from the RBC/UKQCD collaboration with lattice spacing $a^{-1}\unit{=1.78(4)}{\giga\eV}$.}

\FullConference{31st International Symposium on Lattice Field Theory - LATTICE 2013\\
		July 29 - August 3, 2013\\
		Mainz, Germany}

\begin{document}

\section{Introduction}
Lattice simulations in particle physics have traditionally been performed in the isospin limit (i.e. with $m_u = m_d$) and without QED (i.e. the charge is zero for all quarks). Now that the precision of some of the results is approaching numbers of the order of $1\%$, the isospin breaking effects must be included. The current values quoted by the Particle Data Group \cite{Beringer:1900zz} are summarised as

\begin{eqnarray}
\nonumber m_u &=& \; \unit{2.3^{+0.7}_{-0.5}}{\mega\eV} \;\;\;\; q_u= +\frac{2}{3} e\\
m_d &=& \; \unit{4.8^{+0.7}_{-0.3}}{\mega\eV} \;\;\;\; q_d= -\frac{1}{3} e \;.
\label{eq:upanddownpdg}
\end{eqnarray}

Perhaps the most noticeable way in which the breaking of isospin symmetry manifests itself is in the mass splitting of the pseudoscalar meson octet. For instance, the charged and neutral pions, which are made of up and down quarks would have the same mass if isospin were conserved. Experimental measurements show a $\unit{4.6}{\mega \eV}$ difference which is about $3\%$ of the mass of the neutral pion.

The baryon sector shows a similar difference. Protons and neutrons, which are also made from up and down quarks, have a $\unit{1.3}{\mega\eV}$ mass difference which is a $0.1\%$ effect \cite{Beringer:1900zz}:

\begin{equation}
m_p - m_n = \unit{-1.2933322(4)}{\mega\eV} \;.
\label{eq:protonneutrondiff}
\end{equation}
This tiny difference, however, is very important as it plays a role in $\beta$-decay and nuclear stability since the sign of the mass difference makes the proton and the hydrogen atom stable.
The expected size of isospin breaking effects is roughly 1\% . Since we are in an era of lattice calculations that probe this level of accuracy, it is necessary to include effects of this order. We need to be able to accurately predict the experimental results to confirm that the theory of QCD and ultimately the Standard Model is correct. The main goal of this work is to find the non-degenerate up, down and strange quark masses. For a review of the current status of calculations see \cite{Portelli:2013jla} and \cite{Colangelo:2010et}. This work builds upon \cite{Duncan:1996xy} (historical), \cite{Blum:2007cy} and \cite{Blum:2010ym}.

\section{Details of Simulation}

\begin{table}
    \begin{tabular}{llllllll}
      Action & $\beta$ & $(am_{ud,sea})$ & $(am_{val})$ &$(am_{s,sea})$ & L/a & $N_{conf}$ & $a$ (fm) \\ \hline
      I & 2.13 & 0.005 & \scriptsize{0.001, 0.005, 0.01, 0.02, 0.03} & 0.04  & 24 & 195 & 0.114 \\
      ~ & ~ & 0.01 & \scriptsize{0.001, 0.01, 0.02, 0.03} & ~  & ~ & 180 & ~ \\
      ~ & ~ & 0.02 & \scriptsize{0.02} & ~  & ~ & 360 & ~ \\
      ~ & ~ & 0.03 & \scriptsize{0.03} & ~  & ~ & 360 & ~ \\ \hline
      I & 2.25 & 0.04 & \scriptsize{0.04} & 0.03  & 32 & 131 & 0.086 \\
      ~ & ~ & 0.06 & \scriptsize{0.06} & ~  & ~ & 188 & ~ \\
      ~ & ~ & 0.08 & \scriptsize{0.08} & ~  & ~ & 81 & ~ \\ \hline
      I+DSDR & 1.75 & 0.001 & \scriptsize{0.001} & 0.045  & 32 & 13 & 0.146 \\
      ~ & ~ & 0.0042 & \scriptsize{0.0042} & ~  & ~ & 12 & ~ \\
    \end{tabular}
    \label{tab:ensembles}
    \caption{Summary of ensembles used in present work. Gauge actions are Iwasaki(I) and Iwasaki+DSDR (I+DSDR).}
\end{table}

The calculations are carried out on QCD ensembles generated by the RBC and UKQCD collaborations, using domain wall fermions (DWF) with 2+1 flavours and Iwasaki gauge action. Previous work \cite{Blum:2010ym} has relied on two lattice sizes: $16^3 \times 64 \times 16$ and $24^3 \times 64 \times 16$, which have volume $\unit{1.8}{\fm^3}$ and $\unit{2.7}{\fm^3}$ respectively with gauge coupling $\beta = 2.13$ and lattice cutoff $\unit{1.78}{\giga\eV}$. This work will extend the study to $32^3\times 64$ Iwasaki, and will additionally use the Iwasaki+DSDR (Dislocation Suppressing Determinant Ratio) gauge action \cite{Arthur:2012opa}. This has lattice cutoff $\unit{2.28(3)}{\giga\eV}$ for Iwasaki and $\unit{1.37(1)}{\giga\eV}$ for Iwasaki+DSDR. For Iwasaki, there are $3$ bare light masses of $0.004$, $0.006$ and $0.008$ combined with one strange of $0.03$. The lightest unitary pion for Iwasaki is $\unit{293}{\mega\eV}$. For Iwasaki+DSDR, the light masses are $0.001$ and $0.0042$ with a strange of $0.045$, which is close to physical. The lightest unitary pion for this lattice is $\unit{170}{\mega\eV}$. For each combination of masses, the charges of each of the quarks are varied from values in the set $\{-2, -1, 0, 1, 2\}$ in units of $e/3$. 

\section{Background Theory}
The direct simulation of QED on the lattice poses some technical difficulties. The non-compact formulation of QED has the action

\begin{eqnarray}
S_{QED}[A] = \frac{1}{4} \int d^4 x \; (\partial_\mu A_\nu(x) - \partial_\nu A_\mu(x))^2 \;.
\label{eq:qedaction}
\end{eqnarray}
This avoids photon self-interactions that the compact formulation would encounter. The action in Eq. (\ref{eq:qedaction}) describes a free theory and since the action is Gaussian, it is trivial to generate the configurations. The generated QED field $A_\mu^{QED}$ is promoted to a compact link variable by exponentiating it. It is then coupled to QCD by multiplying it by the QCD gauge link $U^{QCD}_\mu$,

\begin{eqnarray}
U_\mu = \exp(ieQA^{QED}_\mu)U^{QCD}_\mu\;.
\label{eq:compactifyqed}
\end{eqnarray}
In quenched QED, the sea quarks are electromagnetically neutral and couple only to $U^{QCD}_\mu$. For this reason, the QCD gauge links are not generated with the QED action. As a consequence, the QCD configurations that were previously generated by the RBC/UKQCD collaboration can be reused.

A problem with the photon is immediately encountered. Since it is massless, it propagates a long distance and finite volume effects may be expected to be large. To counter this, finite volume correction terms are included in the fits \citep{Blum:2010ym,Hayakawa:2008an}. Another problem comes when the photon propagator is naively discretised. Refs. \cite{Duncan:1996xy, Blum:2007cy} estimate the QED finite size scaling by the one-pole saturation approximation with a momentum integral replaced by a sum. The pion mass-squared difference in the effective theory including vector and axial mesons then encounters terms such as

\begin{eqnarray}
\int \frac{dk^0}{2\pi}\frac{1}{L^3} \sum_{\mathbf{k} \in \tilde\Gamma_3}\frac{m_V^2 m_A^2}{k^2(k^2+m^2_V)(k^2+m^2_A)}
\label{eq:discretiseintegral}
\end{eqnarray}
where $\tilde\Gamma_3 \equiv \lbrace \mathbf{k} = (k^1, k^2, k^3) \; \vert \;k^j \in \frac{2 \pi}{L}\mathbb{Z} \rbrace$, $L$ is the length of the lattice in the spatial direction, $m_V \simeq \unit{770}{\mega\eV}$ and $m_A \simeq \unit{970}{\mega\eV}$ is the chiral limit mass of the $A_1$. The $\mathbf{k}=0$ term of this sum diverges for finite volume. We notice that any single point of the integral has measure zero, so the value of the integral is not changed if we change one point. For the sum, we then take the first term of the series to zero to remove the zero mode of the photon propagator. This approach is justified in \cite{Hayakawa:2008an}. In this work the Feynman gauge is used to gauge fix non-compact QED. For another discussion see {\cite{Portelli:2010yn}.

According to Dashen's theorem, the lowest order EM effect, the Dashen term entering at $\bigo{(\alpha_{EM})}$, is the dominant contribution to the charged-neutral pion mass difference. As the chiral limit is approached (i.e. massless quarks), this relation also holds for kaons. This can be summarised by the relation

\begin{equation}
\Delta_{QED} M_K^2 - \Delta_{QED} M_\pi^2 = 0 \;,
\label{eq:dashen}
\end{equation}
where $\Delta_{QED} M_P^2 = (M_{P^\pm}^2 - M_{P^0}^2)_{m_u = m_d}$ for $P=\pi, K$. The relation in Eq. (\ref{eq:dashen}) is violated by $\bigo{(\alpha_{EM} m)}$ terms away from the chiral limit. Using chiral perturbation theory \cite{Bijnens:2006mk}, these corrections can be identified. It is then possible to find the non-degenerate quark masses by matching to experimentally measured mass splittings. These violations to Dashen's theorem can be parametrised by the FLAG parametrisation \cite{Juttner:2011jg}

\begin{eqnarray}
\varepsilon = \frac{\Delta_{QED} M_K^2 - \Delta_{QED} M_\pi^2}{\Delta M_\pi^2} \;.
\end{eqnarray}

Since the simulations have unnaturally large up and down masses, the pions will have a larger mass too. It is therefore necessary to know how the masses of the pion and kaon change as a function of the input masses. Chiral perturbation theory can be used to find the mass-squared of the pion and kaon as a function of the quark masses and charges. As a consequence of doing this, low energy constants (LECs) are also necessary, which are non-perturbative coefficients to the various terms in the equations. This work uses $\mbox{SU}(2) \times \mbox{SU}(2)$ chiral perturbation theory for the pions including photons to next to leading order (NLO) including the kaons \cite{Blum:2010ym}. This includes one-loop logs proportional to $\alpha_{em} m$. It was found in \cite{Allton:2008pn} that $\mbox{SU}(3)$ was poorly convergent for masses close to the physical strange mass, making it inappropriate for finding physical results.

The mass-squared of the pion and kaon are functions of the input masses and charges \cite{Blum:2010ym}. The mass-squared formulae also depend on the QCD and QED low energy constants, which will need to be fit and calculated. For example, the (infinite volume) mass-squared of the pion as a function of the masses and charges is
\begin{align}
\label{eq:chptpion}
M^2 &= \chi_{13}\left\{1+\frac{24}{{\color{blue}F^2}}({\color{blue}2 L_6-L_4})\frac{\chi_4+\chi_{5}}{3}
+\frac{8}{{\color{blue}F^2}}({\color{blue}2 L_8-L_5})\chi_{13}\right.\nonumber\\
&\quad\quad
+
\left.
 \frac{1}{2}\,\frac{1}{16\pi^2 {\color{blue}F^2}}
 \left(
    R^\pi_{13}\, \chi_\pi\,\log \frac{\chi_\pi}{\mu^2}
  + R^1_{\pi 3}\,\chi_1\,\log \frac{\chi_1}{\mu^2}
  + R^3_{\pi 1}\,\chi_3\,\log \frac{\chi_3}{^2}
 \right)
\right\}\nonumber\\
&-12e^2  {\color{red}Y_1}  \bar q^2 \chi_{13} +4e^2  {\color{red}Y_2} q_p^2 \chi_{p} +4 e^2  {\color{red}Y_3} q_{13}^2 \chi_{13}
- 4 e^2  {\color{red}Y_4} q_{1}q_{3} \chi_{13} +12e^2  {\color{red}Y_5} q_{13}^2 \frac{\chi_{4}+\chi_5}{3}\nonumber\\
&- e^2 \frac{3}{16\pi^2}\chi_{13}\log{\frac{\chi_{13}}{\mu^2}}q_{13}^2+e^2 \frac{1}{4\pi^2}\chi_{13}q_{13}^2 +\frac{2{\color{red}C}e^2}{{\color{blue}F^2}}q_{13}^2 +
e^{2}{\color{red}\delta_{m_{\rm res}}}(q_1^2+q_3^2)
\nonumber\\
&-e^2 \frac{{\color{red}C}}{{\color{blue}F^4}} \frac{1}{8\pi^2}q_{13}\left(q_{14}
\chi_{14}\log{\frac{\chi_{14}}{\mu^2}}+q_{15}\chi_{15}\log{\frac{\chi_{15}}{\mu^2}}
-q_{34}\chi_{34}\log{\frac{\chi_{34}}{\mu^2}}-q_{35}\chi_{35}\log{\frac{\chi_{35}}{\mu^2}}
\right) \; ,
\end{align}
where $\chi_{ij} = {\color{blue}B}(\tilde{m_i} + \tilde{m_j})$, $\tilde{m_i} = m_i + m_{res}$ where $m_i$ is the bare quark mass and $m_{res}$ is the residual mass. The terms $m_{res}$ and $\delta_{m_{\rm res}}$ are lattice artifacts arising from the finite extent of the 5th dimension in the formulation of DWF. The definitions of all the terms can be found in \cite{Blum:2010ym}. There are $6$ QCD ($\color{blue} B, F, L_4, L_5, L_6, L_8$)  and $6$ QED LECs ($\color{red} C, Y_1, Y_2, Y_3, Y_4, Y_5, \delta_{mres}$) to fit. These are defined at the scale $\mu=\unit{1}{\giga\eV}$ and are highlighted in the equation. The coefficient $Y_1$ is included in the formula for completion, but since this work is done in quenched QED it cannot be determined. The effect of neglecting this is included in the estimation of the error in \cite{Blum:2010ym}.

\section{Strategy}
In \cite{Blum:2010ym}, values for the non-degenerate quark masses were found with the $16^3$ and $24^3$ lattices. This work extends these results with finer $32^3$ ensembles. This will allow a continuum extrapolation to be performed while exploiting the finer lattices for improved precision.
To find the non-degenerate light and strange quark masses, the formulae for the mass-squared of the pion and kaon are matched to experimental values. The charged and neutral kaon and the charged pion are used and the matching is performed by varying the quark masses simultaneously to minimise the $\chi^2$ value of the fit. 
To do this, charged and neutral mesons are simulated and their masses are found in the standard way of fitting to exponentials. The mass-squared difference of the charged and neutral meson are then fit to find the QED LECs first. Since there are many different combinations of the masses and charges and corresponding mass-squared differences, it is possible to get an estimate of these LECs by performing best fits of the parameters. The non-degenerate quark masses are then found by matching the mass-squared formulae to the experimental meson masses by varying the quark masses simultaneously. The neutral pion is not used to fix the parameters, since this would involve calculating disconnected diagrams, which have a worse signal to noise ratio than connected diagrams. Errors on the masses used in the fits are found using the standard jackknife procedure.

\section{Results}
The non-degenerate up and down quark masses were found by setting $m_s = \unit{95}{\mega\eV}$ and fitting to the functional forms of $M^2_{K^+}$, $M^2_{K^0}$ and $M^2_{\pi^+}$. Even though there are two fit parameters and three target quantities naively, the fits are performed with the squared difference of the kaons $M^2_{K^+} - M^2_{K^0}$ and $M^2_{\pi}$, reducing the number of quantities by one. This is done because it was found that $M^2_{K^+}$ and $M^2_{K^0}$ individually had poor sensitivity to the up and down quark masses but the combination $M^2_{K^+} - M^2_{K^0}$ was more sensitive. The results are summarised in Tab. \ref{tab:results} for the infinite and finite volume fits. All results are preliminary. Numbers are quoted in $\overline{\mbox{MS}}$ at $\unit{2}{\giga\eV}$.
To also have $m_s$ as a fit parameter, an extra strange quark mass is necessary which is added by using the Iwasaki+DSDR ensembles. Work is continuing to increase the statistics on these ensembles.

\begin{table}
    \begin{tabular}{lllll}
      ~ & $m_u$ (MeV)&$m_d$ (MeV)&$m_u / m_d$&$m_d - m_u$ (MeV) \\ \hline
      Infinite Volume & $2.264(82)$ & $4.815(40)$ & $0.471(18)$ & $2.551(97)$ \\
      Finite Volume & $2.137(85)$ & $4.680(45)$ & $0.457(19)$ & $2.543(98)$ \\
    \end{tabular}
    \caption{Preliminary results for $32^3\times64\times16$ Iwasaki ensembles.}
    \label{tab:results}
\end{table}

\begin{figure}[H]
    \centering
        \input{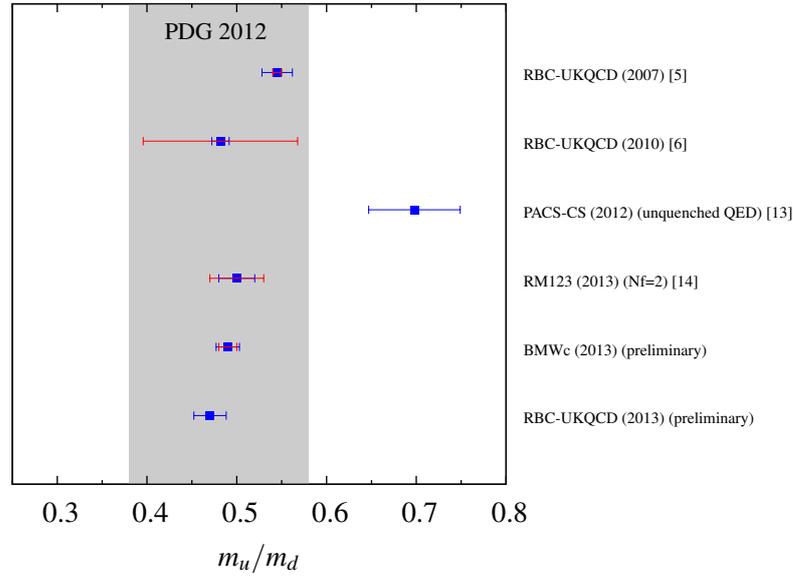}
        \vspace{0.1cm}
    \caption{\scriptsize{Summary of lattice calculations of the up and down quark mass ratio. Source: \cite{Portelli:2013jla}}}
    \label{fig:mqlat}
\end{figure}

\section{Conclusions \& Outlook}
These results show that it is possible to discriminate small isospin breaking effects in state of the art lattice simulations. To gain a better understanding it is important to do a global fit of all of the ensembles that the RBC/UKQCD collaboration have generated. To complete this, efforts are being made to increase the statistics of the Iwasaki+DSDR ensembles.

\end{document}